\documentclass[fleqn,10pt]{wlscirep}

\usepackage{setspace}
\title{No-cloning of quantum steering}

\author[1]{Ching-Yi Chiu}
\author[2]{Neill Lambert}
\author[1]{Teh-Lu Liao}
\author[2,3]{Franco Nori}
\author[1,*]{Che-Ming Li}

\affil[1]{Department of Engineering Science, National Cheng Kung University, Tainan 701, Taiwan}
\affil[2]{CEMS, RIKEN, Wako-shi, Saitama 351-0198, Japan}
\affil[3]{Department of Physics, University of Michigan, Ann Arbor, Michigan 48109-1040 USA}

\affil[*]{cmli@email.ncku.edu.tw}

\begin{abstract}
Einstein-Podolsky-Rosen (EPR) steering allows two parties to verify their entanglement, even if one party's measurements are untrusted. This concept has not only provided new insights into the nature of non-local spatial correlations in quantum mechanics, but also serves as a resource for one-sided device-independent quantum information tasks. Here, we investigate how EPR steering behaves when one-half of a maximally-entangled pair of qudits (multidimensional quantum systems) is cloned by a universal cloning machine. We find that EPR steering, as verified by a criterion based on the mutual information between qudits, can only be found in one of the copy subsystems but not both. We prove that this is also true for the single-system analogue of EPR steering. We find that this restriction, which we term ``no-cloning of quantum steering'', elucidates the physical reason why steering can be used to secure sources and channels against cloning-based attacks when implementing quantum communication and quantum computation protocols.
\end{abstract}

\begin{document}
\maketitle

\section*{Introduction}

Einstein-Podolsky-Rosen (EPR) steering reveals that one party, Alice, can affect, or steer, another remote party (Bob's) state, by her measurements on one particle of an entangled pair shared between them \cite{Schrodinger35}. This concept was originally introduced by Schr\"{o}dinger in response to the EPR paradox \cite{EPR35}. Recently, it has been reformulated by Wiseman, Jones and Doherty \cite{Wiseman07} as a information-theoretic task to demonstrate that Alice and Bob
can validate shared entanglement even if the measurement devices of Alice are untrusted. This has led to a range of conceptually important extensions of the concept of EPR steering and several potential applications for practical quantum information processing. See an in-depth discussion given in the review by Reid \textit{et al.} \cite{Reid09}.

As articulated by Wootters and Zurek \cite{Wootters82} and Dieks \cite{Dieks82} in 1982, it is impossible to perfectly copy an unknown quantum state. This famous no-go theorem of quantum mechanics has significant implications in understanding nonclassical features of quantum systems and profound applications in quantum information science. Although one cannot make perfect copies of an unknown quantum state, it is possible to create imperfect copies. Bu\v{z}ek and Hillery \cite{Buzek96} have shown that a universal cloning machine can produce a clone of an unknown state with high fidelity. Such a universal cloning machine has been shown to be optimal and has been extensively studied in the context of possible alternatives, extensions and use as an eavesdropping attack on the protocols of quantum cryptography \cite{Scarani05}.

\begin{figure*}[t]
\includegraphics[width=17 cm,angle=0]{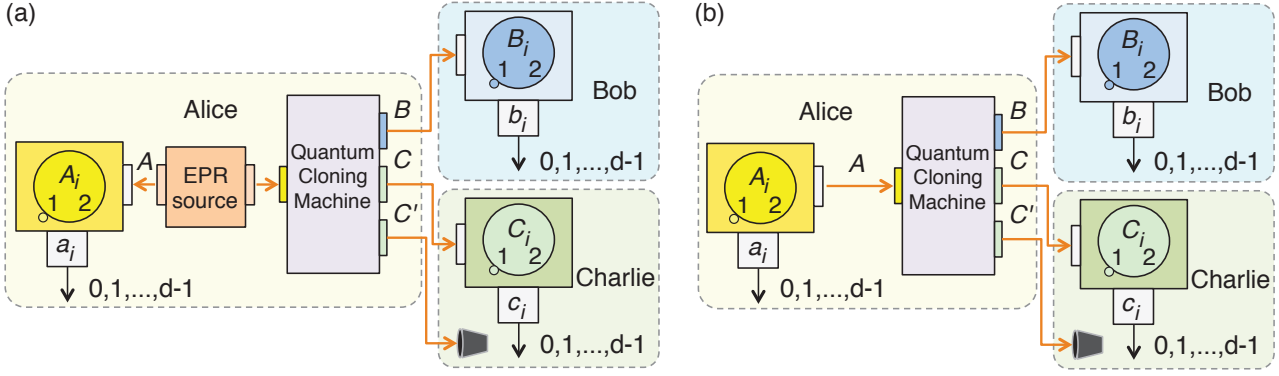}
\caption{Cloning quantum steering. (a) Einstein-Podolsky-Rosen (EPR) steering. Alice creates maximally entangled pairs $\left|\Phi\right\rangle$ (\ref{phi}) from an EPR source. She keeps one qudit ($A$) and sends the other qudit of the pair into a universal cloning machine. The cloning machine, assisted by ancilla qudits (not shown), creates a four-partite composite state (\ref{phiabcc}). After cloning, the qudit $B$ is sent to Bob and the qudit $C$, together with the ancilla $C'$, are sent to Charlie. Each of the three parties has an apparatus to implement two complementary measurements $m_{i}$ for $m=A,B,C$ and $i=1,2$. Their measurement results $n_{i}\in\{0,1,...,d-1\}$ for $n=a,b,c$ are then used to certify EPR steering of the subsystems $(A,B)$ and $(A,C)$ using a steering criteria~(\ref{entro0}). (b) Single-system (SS) steering: A qudit with the state $\left|s\right\rangle_{Ai}$ is sent from Alice to a cloning machine. Here $\left|s\right\rangle_{Ai}$ is a post-measurement state of some initial qudit (not shown) under the measurement $A_{i}$ for $i=1,2$. A tripartite composite system is then created by the cloning machine, and the qudit $B$ is sent to Bob and the qudit $C$, together with the ancilla $C'$, are sent to Charlie. The measurement apparatus used by each party are the same as the devices used in the case of EPR steering. They can also use a steering criterion (\ref{entro0}) to identify the SS steering of the subsystems $(A,B)$ and $(A,C)$.}\label{Quantumcloning}
\end{figure*}

Here, inspired by the no-cloning theorem and the concept of quantum steering, we ask a simple question: "Does quantum mechanics allow quantum steering to be copied by a universal cloning machine?". To investigate this question, we use the concept of a universal cloning machine to consider how quantum steering is cloned and shared between two copies of a qudit (a multidimensional quantum  system) which itself is half of a maximally-entangled pair [see Fig.~\ref{Quantumcloning}(a)]. In addition, we apply the same method of analysis to the single-system (SS) analogue of EPR steering (SS steering) scenario \cite{Li15} [Fig.~\ref{Quantumcloning}(b)]. We find that EPR steering (and SS steering), as described by a criterion based on the mutual information between two parties, can only be observed in one of the two copy subsystems, but not both. We denote this as the ``no-cloning of steering". Several applications to quantum information directly follow, such as (i) the observation of steering validates channels against cloning-based coherent attacks when implementing quantum key distribution (QKD) and (ii) steerability guarantees the reliability of quantum logic gates of arbitrary size for both the quantum circuit model and one-way quantum computing.  They give physical insight into the observation in earlier works that various steering criteria vanish when the noise in a channel passes the threshold for secure QKD and quantum computation\cite{Li15}.

\section*{Results}

\textbf{Quantum steering and steering criteria}

\noindent EPR steering typically consists of two steps: First, Alice generates a bipartite entangled system from an entanglement source [often called an EPR source, see Fig.~\ref{Quantumcloning}(a)]. To have a concrete illustration, let us assume that this entangled state is of the form
\begin{equation}
\left|\Phi\right\rangle=\frac{1}{\sqrt{d}}\sum_{s=0}^{d-1}\left|s\right\rangle_{A1}\otimes\left|s\right\rangle_{B1}\label{phi}
\end{equation}
where $\left|s\right\rangle_{A1}=\left|s\right\rangle_{B1}=\left|s\right\rangle_{1}$, where $\{\left|s\right\rangle_{1}|s=0,1,...,d-1\}$ is an orthonormal basis that corresponds to bases of Alice's measurement $A_{1}$ and Bob's measurement $B_{1}$. Second, Alice keeps one qudit of the entangled pair and sends the other qudit to Bob. Then, depending on Alice's measurement result $a_{1}=s$, the state of the qudit finally held by Bob can be steered into a corresponding quantum state, $\left|s\right\rangle_{B1}$, for the result $b_{1}=s$. Such remote preparation of Bob's states can be also be seen in other bases. For example, suppose that Alice and Bob's measurements $A_{2}$ and $B_{2}$ correspond to another orthonormal basis $\{\left|s\right\rangle_{2}|s=0,1,...,d-1\}$, where $\left|s\right\rangle_{2} =1/\sqrt{d
}\sum_{k=0}^{d-1}\exp (i \frac{2\pi}{d} s k)\left| k\right\rangle_{1}$, the state vector of $\left|\Phi\right\rangle$ represented in this basis is of the form $\left|\Phi\right\rangle=1/\sqrt{d}\sum_{s=0}^{d-1}\left|s\right\rangle_{A2}\otimes\left|d-s\right\rangle_{B2}$,
where $\left|s\right\rangle_{A2}=\left|s\right\rangle_{B2}=\left|s\right\rangle_{2}$. It is clear that Bob's outcome $b_{2}$ will respond to Alice's outcome $a_{2}$, which satisfies $a_{2}+b_{2}\doteq 0$, where $\doteq$ denotes equal module $d$. Such dependence can be made manifest by the conditional entropy $H(B_{1}|A_{1})=H(B_{2}|A_{2})=0$, where $H(B_{i}|A_{i})\equiv-\sum_{a_{i}=0}^{d-1}P(a_{i})\sum_{b_{i}=0}^{d-1}P(b_{i}|a_{i})\log_{2}P(b_{i}|a_{i})$. In practical experiments, the marginal probabilities $P(a_{i})$ and the conditional probabilities $P(b_{i}|a_{i})$ can, in principle, be measured to explicitly consider this dependence.

This description of EPR steering can be directly mapped to single-system or temporal steering and vice-versa (see \cite{Li15} for detailed discussions). As depicted in Fig.~\ref{Quantumcloning}(b), first, Alice prepares a qudit with the state $\left|s\right\rangle_{Ai}$ by performing complementary measurements $A_{1}$ or $A_{2}$ on an initial state. Second, Alice sends the prepared qudit to Bob. Then she can steer the state Bob holds $\left|s\right\rangle_{Bi}$ ($\left|s\right\rangle_{Bi}=\left|s\right\rangle_{Ai}$ for the ideal case) into other quantum state by, for example, asking Bob, via a classical channel, to perform a unitary transformation on  $\left|s\right\rangle_{Bi}$.

In practical situations, demonstrations of both EPR steering and SS steering are imperfect. Environmental noise, or randomness introduced by an eavesdropper, can affect both the quantum source for creating $\left|\Phi\right\rangle$ and $\left|s\right\rangle_{Ai}$ and the properties of the state during its transmission from Alice to Bob. In addition, in its information task formulation, Bob also does not trust Alice nor her measurement apparatus, and wishes to verify whether she is truly steering his state. Hence, it is important to have an objective tool that can certify the ability of Alice to steer the states of the particles eventually held by Bob. Here, we describe and verify quantum steering in terms of the mutual information between measurement results of Alice and Bob $I_{A_{i}B_{i}}=H(B_{i})-H(B_{i}|A_{i})$. Earlier works showed that if the mutual dependence between Alice and Bob's measurement results violates the bound \cite{Li15}
\begin{equation}
\sum_{i=1}^{2}I_{A_{i}B_{i}}>\log_{2}d,\label{entro0}
\end{equation}
their dependence is stronger than the correlation between Bob's outcomes and the results derived from unsteerable states alone, verifying Alice's ability to steer Bob's state. As shown in \cite{Li15}, it is worth noting that the entropic steering criteria (\ref{entro0}) are applicable to both EPR steering and SS steering.  One difference between them is that $P(b_{i}|a_{i})$ for SS steering are derived from measurements on single systems where $a_{i}$ and $b_{i}$ are taken at two different times.\\

\noindent\textbf{No-cloning of quantum steering}

\noindent Suppose that Alice has an entanglement source to create pairs of qudits $\left|\Phi\right\rangle$. One qudit of the entangled pair is sent to a universal cloning machine and the other qudit ($A$) is kept by Alice. See Fig.~\ref{Quantumcloning}(a). After passing through the cloning machine, two new qudits are created, and the state of the total system becomes
\begin{equation}
\left|\phi\right\rangle_{ABCC'}=\sum_{j,k=0}^{d-1}\sqrt{\lambda_{jk}}\left|\phi_{jk}\right\rangle_{AB}\left|\phi_{j,d-k}\right\rangle_{CC'}.\label{phiabcc}
\end{equation}
The qudit $B$ is sent to Bob whereas the qudits $C$ and $C'$ are sent to a third party Charlie. The two-qudit state vectors $\left|\phi_{jk}\right\rangle_{AB}$ and $\left|\phi_{j,d-k}\right\rangle_{CC'}$ are described by
\begin{eqnarray}
\left|\phi_{jk}\right\rangle_{mn}&=&(I\otimes U_{j,k})\left| \Phi\right\rangle\nonumber\\
&=&\frac{1}{\sqrt{d}}\sum_{s=0}^{d-1}\exp (i \frac{2\pi}{d} s k)\left| s\right\rangle_{m1}\left| s+j\right\rangle_{n1},
\end{eqnarray}
for $(m,n)=(A,B),(C,C')$, where $I$ denotes the identity operator, $U_{j,k}=\sum_{s=0}^{d-1}\exp(i2\pi sk/d)\left|s+j\right\rangle_{n1n1}\left\langle s\right|$, and $\left| s\right\rangle_{m1}=\left| s\right\rangle_{n1}=\left|s\right\rangle_{1}$. The state of Alice's and Bob's qudits is
\begin{equation}
\rho_{AB}=\sum_{j,k=0}^{d-1}\lambda_{jk}\left|\phi_{jk}\right\rangle_{AB AB}\left\langle \phi_{jk}\right|,
\end{equation}
where $\lambda_{jk}$ denotes the probability of observing $\left|\phi_{jk}\right\rangle_{AB}$. The mutual information of Alice's and Bob's measurement results derived from their measurements $A_{i}$ and $B_{i}$ on $\rho_{AB}$ is

\begin{equation}
I_{A_{i}B_{i}}=\log_{2}d-\sum_{t=0}^{d-1}q_{i}^{t}\log_{2}\frac{1}{q_{i}^{t}},\label{Iab}
\end{equation}
where $q_{1}^{t}=\sum_{k=0}^{d-1}\lambda_{tk}$ and $q_{2}^{t}=\sum_{j=0}^{d-1}\lambda_{j,d-t}$. The variables $q_{i}^{t}$ firstly introduced in \cite{Sheridan10} are the probabilities of finding $b_{i}-a_{i}=t$ or $b_{i}-a_{i}=t-d$ for $t=0,1,..,d-1$. The sum of mutual information under two measurement settings is then
\begin{equation}
\sum_{i=1}^{2}I_{A_{i}B_{i}}=2\log_{2}d-\sum_{i=1}^{2}H(q_{i}^{t}).\label{iabfinal}
\end{equation}

To determine the mutual information of Alice's and Charlie's measurement results $I_{A_{i}C_{i}}$, we first consider the mutual dependence between $a_{i}$ and the results derived from measurements on the subsystem composed of Charlie's qudit $C$ and the ancilla $C'$ by their mutual information $I_{A_{i}(C_{i}C'_{i})}$. It is clear that
\begin{equation}
I_{A_{i}C_{i}}\leq I_{A_{i}(C_{i}C'_{i})}.\label{iacacc}
\end{equation}
In addition, the mutual information $I_{A_{i}(C_{i}C'_{i})}$ is constrained by the Holevo bound by
\begin{equation}
 I_{A_{i}(C_{i}C'_{i})}\leq S(\rho_{CC'})-\sum_{a_{i}=0}^{d-1}P(a_{i})S(\rho_{CC'|a_{i}}).\label{Hbound}
\end{equation}
$S(\rho_{CC'})$ is the von-Neumann entropy of the state $\rho_{CC'}=\sum_{j,k=0}^{d-1}\lambda_{jk}\left|\phi_{j,d-k}\right\rangle_{CC'CC'}\left\langle \phi_{j,d-k}\right|$. It can be explicitly represented by
\begin{equation}
S(\rho_{CC'})=\sum_{j,k=0}^{d-1}-\lambda_{jk}\log_{2}\lambda_{jk}\equiv H(\lambda).
\end{equation}
The state $\rho_{CC'|a_{i}}$ is the reduced state conditioned when Alice obtains the result $a_{i}$. Now, we use the method presented in \cite{Sheridan10} to find the upper bound in (\ref{Hbound}). The von-Neumann entropy of this state can be shown as $S(\rho_{CC'|a_{i}})=\sum_{t=0}^{d-1}q_{i}^{t}\log_{2}\frac{1}{q_{i}^{t}}\equiv H(q_{i}^{t})$. In order to derive the upper bound of $ I_{A_{i}(C_{i}C'_{i})}$ by minimizing the difference between $S(\rho_{CC'})$ and $\sum_{a_{i}=0}^{d-1}P(a_{i})S(\rho_{CC'|a_{i}})$, we substitute $\lambda_{j,d-k}=f(j,k)q_{1}^{j}$ into $q_{2}^{t}=\sum_{j=0}^{d-1}\lambda_{j,d-t}$, where $\sum_{k=0}^{d-1}f(j,k)=1$, and then obtain  $q_{2}^{t}=\sum_{k=0}^{d-1}f(t,k)q_{1}^{k}$. For each $t$, all $f(t,k)=q_{2}^{t}$ implies the minimum of the difference. Then we have
\begin{equation}
 H(\lambda)=H(q_{1}^{t})+\sum_{t}q_{1}^{t}H(f(t))=H(q_{1}^{t})+H(q_{2}^{t}).\label{hq12}
\end{equation}
With Eqs.~(\ref{iacacc}), (\ref{Hbound}) and (\ref{hq12}), the upper bound of the mutual information $I_{A_{i}C_{i}}$ is shown as
\begin{equation}
I_{A_{i}C_{i}}\leq H(q_{1}^{t})+H(q_{2}^{t})-H(q_{i}^{t}),
\end{equation}
which implies that
\begin{equation}
\sum_{i=1}^{2}I_{A_{i}C_{i}}\leq \sum_{i=1}^{2}H(q_{i}^{t}).\label{iactemp}
\end{equation}
Hence, combining Eq.~(\ref{iabfinal}) with Eq.~(\ref{iactemp}), we eventually derive the following relationship between the mutual information of Bob's and Charlie's systems with Alice's
\begin{equation}
\sum_{i=1}^{2}I_{A_{i}C_{i}}+\sum_{i=1}^{2}I_{A_{i}B_{i}}\leq 2\log_{2}d.\label{iacfinal}
\end{equation}
This criterion~(\ref{iacfinal}) provides a basis to investigate how EPR steering is shared between two copies of a qudit of a maximally entangled pair. When the correlation between the qudits shared by Alice and one of the two parties, say Charlie, is certified by the steering criteria (\ref{entro0}), it is clear that the mutual dependence between qudits shared by Alice and Bob will not be stronger than an unsteerable state. Hence, EPR steering can be identified in only one of the copy subsystems. This analysis of the behavior of EPR steering subject to cloning can be directly applied to SS steering as well; see Methods section.\\

\noindent\textbf{Securing quantum information processing}

\noindent The steerability of Alice over Bob or Charlie's qudits, as certified by the steering criteria~(\ref{entro0}), implies that the mutual dependence between them is stronger than the mimicry that an unsteerable state can provide. In addition, such steering cannot be shared with a third party by using a universal cloning machine. Two direct applications to quantum information are illustrated as follows.

(i) If a sender (Alice) and a receiver (Bob) confirm that their measurement results are classified as steerable, according to the criteria (\ref{entro0}), they can be convinced that an eavesdropper (Charlie) who uses a cloning machine for coherent attacks cannot produce states that can be steered by the sender. This is because the mutual information between Alice and Bob is larger than the mutual information shared between Alice and the eavesdropper, Charlie. Thus they can use privacy-amplification techniques on their shared measurement outcomes to generate a secure key. Thus the no-cloning of quantum steering verified by (\ref{entro0}) shows that ruling out false steering secures channels against cloning-based attacks when implementing QKD.

(ii) As shown in \cite{Li15} , steering quantum systems is equivalent to performing quantum computation. No-cloning of steering provides a strict proof to show that the observation of quantum steering guarantees faithful implementation of a quantum computing implementation in the presence of uncharacteristic measurements and cloning-based attacks.

\section*{Discussion}

We investigated how quantum steering is cloned by a universal cloning machine and shared between two copy subsystems. We showed that it is impossible to observe quantum steering, as described by the mutual information criterion (\ref{entro0}), in the two copies at the same time. This no-cloning of quantum steering ensures secure QKD and faithful quantum gate operations of arbitrary computing size against cloning-based attacks. Our results motivate several open questions. Is the no-cloning of quantum steering applicable to the situation of genuine multipartite multidimensional EPR steering? If it is the case, then such high-order steering would serve as a source for reliable multipartite quantum information processing such as quantum secret sharing. In addition to high-order steering, does one-way steering possess this feature of no-cloning? Finally, if we use a steering measure instead of an entropic criteria (\ref{entro0}), could the partial power of quantum steering in terms of the units of a steering quantifier be copied by the cloning machine? Could the total quantity of steering be conserved after cloning?

Finally, it is interesting to connect our results with other approaches, such as the principle of monogamy of certain quantum correlations \cite{Kay09,Bartkiewicz15}. In particular, the principle of the monogamy of temporal steering, shown in the work \cite{Bartkiewicz15} [see Eq. (5) therein], is consistent with our results, and suggests our criteria can also be interpreted as a monogamy relation in the entropic form. However, whether such a result can provide a relation in the form of Coffman-Kundu-Wootters monogamy inequality [see, for example, Eq. (1) in \cite{Kay09}] still needs further investigation. In addition, He \textit{et al.} \cite{He15} have shown that two-way steering is required to overcome the no-cloning threshold for secure teleportation. This relationship, between no-cloning and EPR steering, also suggests a principle of no-cloning for the correlations utilized for teleportation. Their quantum-information-task-oriented method, to investigate the relationship between the no-cloning theorem and steering, indicates that it may be interesting, in future work, to consider the security threshold for secure quantum teleportation derived from our input-output scenario for cloning quantum steering, and to compare this condition on fidelity with their criterion \cite{He15}.

\noindent\section*{Methods} 

\noindent{\bf No-cloning of SS steering}

\noindent As illustrated in Fig.~\ref{Quantumcloning}(b), after operating the cloning machine on a single system sent from Alice, the state $\left|s\right\rangle_{Ai}$ becomes
\begin{equation}
\left|\phi\right\rangle_{BCC'}=\sum_{j,k=0}^{d-1}\sqrt{\lambda_{jk}}\left|\phi_{jk}\right\rangle_{B}\left|\phi_{j,d-k}\right\rangle_{CC'},
\end{equation}
where
\begin{equation}
\left|\phi_{jk}\right\rangle_{B}=U_{j,k}\left| s\right\rangle_{Bi}.
\end{equation}
(note that $\left|\phi_{00}\right\rangle_{B}=\left|s\right\rangle_{Ai}$). The state of Bob's qudit is then $\rho_{B}=\sum_{j,k=0}^{d-1}\lambda_{jk}\left|\phi_{jk}\right\rangle_{BB}\left\langle \phi_{jk}\right|$. With this reduced state, we obtain the mutual information $I_{A_{i}B_{i}}$ (\ref{Iab}). When considering the mutual information $I_{A_{i}C_{i}}$, it is easy to find that the connection between $A$ and $CC'$ here can be mapped to the case of EPR steering. There are no differences between the states $S(\rho_{CC'|a_{i}})$ together with $S(\rho_{CC'})$ in these two steering cases. Then we arrive again at the result of a constraint on mutual information for subsystems (\ref{iacfinal}). Hence the SS steering can be observed in only one of the copy subsystems.\\

\section*{Acknowledgements}
This work is partially supported by the Ministry of Science and Technology, Taiwan, under Grant Numbers MOST 104-2112-M-006-016-MY3 and No. MOST-104-2221-E-006-132-MY2. This work is partially supported by the RIKEN iTHES Project, the MURI Center for Dynamic Magneto-Optics via the AFOSR award number FA9550-14-1-0040, the IMPACT program of JST and a Grant-in-Aid for Scientific Research (A). N.L. is partially supported by the FY2015 Incentive Research Project. N.L. and F.N. acknowledge the support of a grant from the John Templeton Foundation.

\section*{Author Contributions}
C.-Y.C. and C.-M.L. devised the basic model. C.-Y.C., N.L., T.-L.L. and C.-M.L. established the final framework. N.L., F.N. and C.-M.L. wrote the paper with contributions from all authors.

\section*{Additional Information}
{\bf Competing financial interests:} The authors declare no competing financial interests.

\end{document}